\documentstyle[psfig,rotating]{mn}

\newif\ifAMStwofonts
\AMStwofontstrue

\def\mm{M_{\rm m}}
\def\ms{M_{\rm s}}

\def\gsim{~\rlap{$>$}{\lower 1.0ex\hbox{$\sim$}}}

\def\ltsim{\lower.5ex\hbox{$\; \buildrel < \over \sim \;$}}
\def\gtsim{\lower.5ex\hbox{$\; \buildrel > \over \sim \;$}}
\def\ltsim{\lower.5ex\hbox{$\; \buildrel < \over \sim \;$}}
\def\gtsim{\lower.5ex\hbox{$\; \buildrel > \over \sim \;$}}

\def\vr{{\bf R}}
\def\vF{{\bf F}}
\def\Rv{R_{\rm v}}

\newcommand{\Mpc}{\, {\rm Mpc} }

\def\rhov{{\rho_{_{\rm v}}}}

\include{rotating}

\begin{document}
\textheight 9in
\title[Dynamics of the Bullet]{Back-in-time dynamics of the 
cluster IE 0657-56 (the Bullet System)}

\author[Nusser] { Adi Nusser\\\\
Physics Department and the Asher Space Science Institute- 
Technion, Haifa 32000, Israel\\
{e-mail: adi@physics.technion.ac.il} }

\maketitle

\begin{abstract}
We present a simplified dynamical model of the ``Bullet'' system of two colliding
clusters. 
The model constrains the masses of the system by 
requiring  that  the orbits of 
the main and sub components satisfy the cosmological initial conditions
of vanishing physical separation a Hubble time ago. 
This  is also known as the timing argument. 
The model considers a system embedded in an  over-dense region. 
We argue that a relative speed of $4500 \rm km/s$ between the two components 
is consistent with cosmological conditions
if the system is of a total mass of $2.8\times 10^{15}h^{-1} M_\odot$ is embedded in a region of a (mild) over-density of
10 times the cosmological background density.
Combining this with the lensing measurements of the projected mass, the model yields a ratio
of 3:1 for the mass of the main relative to that of the subcomponent.
The effect of the background weakens  as the relative speed between the 
two components is decreased.   
For relative speeds lower than $\sim 3700\rm km/s$, the timing argument 
yields masses which are too low to be consistent with lensing.
\end{abstract}
 
                           
\section{introduction}

The system IE 0657-56 of two colliding clusters (termed the ``Bullet"),
is particularly interesting among such systems. 
X-ray observations of this system show a prominent bow-shock generated by the
supersonic motion (Mach number of ${\cal M}=3\pm 0.4$) of the sub-cluster (the Bullet) relative to the gaseous component of the
main cluster.  The estimated speed of the Bullet relative to the main cluster is
 $\sim 4700\rm km/s$, while the relative line-of-sight velocities between the groups of galaxies
associated with   the two components is only $600\rm km/s$.
Therefore, the relative motion between the two clusters 
 is almost entirely perpendicular to the line-of-sight (Barrena et al. 2002; 
 Markevitch et al. 2004).

The current dynamical state of the system must be consistent with
that evolved in an expanding Universe. Therefore, all matter belonging 
to the system must originate from a region of vanishing size as we approach the 
initial singularity ($t\rightarrow 0$).   
The positions of objects in the system can be uniquely moved back-in-time 
from their current positions and velocities. 
The masses of these objects can then be tuned so that the cosmological 
constraint at $t\rightarrow 0 $ is satisfied.
For a two component system, this 
 is traditionally known as the timing argument and it has been used 
for studying the system of the Galaxy and M31 (e.g. Kahn \& Woltjer 19969; Peebles 1993). 
Timing arguments have also been applied for the 
 the Bullet system by  
Zhao (2007). This study has not considered 
systems  embedded in an over-dense 
region  of a larger scale. Further,  it differs in the details from the model presented 
here.

 The outline of this paper is as follows. In \S2 we briefly present 
 the lensing measurements and constrain the masses of the main and sub components 
 assuming NFW  forms (see below)  for their density profiles.
 In \S3 we describe our model and present the results for the mass estimates. 
 We conclude in \S4 with a general discussion.

\section{Lensing Constraints}

We assume that the density profiles of dark matter halos
of the main and the sub components 
follow the form given 
by Navarro, Frenk \& White (1996) (hereafter, NFW) as 
\begin{equation}
\rho(r)=\frac{\rho_{\rm crit}\delta_{\rm 0} }{x(1+c x)^2} \; ,
\label{eq:nfw}
\end{equation}
where $c$ is the concentration paramater, $x=r/R_{\rm v}$ is the distance from 
the halo center in units of the virial radius, and $\rho_{\rm crit}=3H^2/(8\pi G)$ is the critical  mean density ($H=\dot a /a$ where $ a$ is the scale factor). 
Here we assume that the mean density  within the
virial radius is $178 \rho_{\rm crit}$. 
This gives $\delta_0$ to be $\delta_0=(178/3) c^2/ [\ln(1+c)-x/(1+c)]$ 

\begin{figure*} 
\centering
\begin{sideways}
\mbox{\psfig{figure=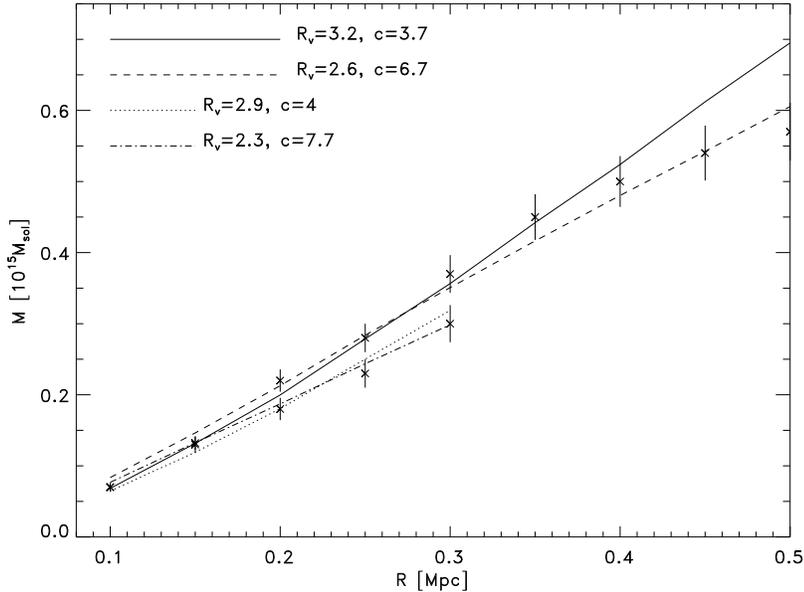,height=4.2in}}
\end{sideways}
\vspace{0.2cm}
\caption{The projected masses of the colliding clusters, as estimated lensing analyses of Brada{\v c} et al. (2006). The upper  points extending to $0.5 \rm Mpc$ show the main halo, while the lower correspond to the sub halo. 
The lines show NFW projected mass profiles.}
\label{fig:lensem}
\end{figure*}

\begin{figure*}
\centering
\begin{sideways}
\mbox{\psfig{figure=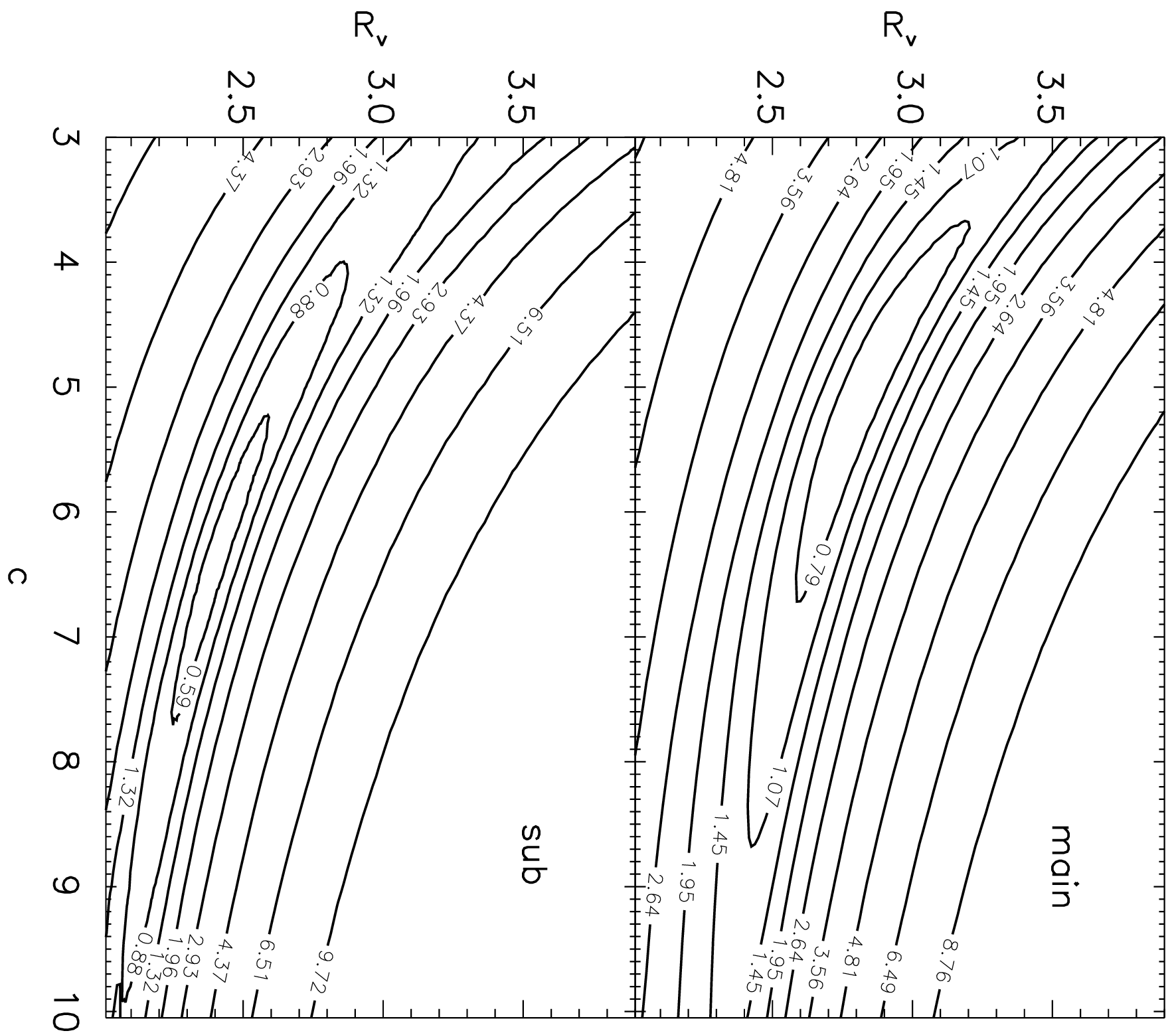,height=4.2in}}
\end{sideways}
\vspace{0.2cm}
\caption{Contours of the quantity $\tilde \chi$
which represents the goodness-of-fit of NFW profiles 
to the lensing data, in the plane of the concentration parameter $c$ and and the virial radius $\Rv$ in units of Mpc.  
(the virial mass is $M_{\rm v}=10^{14}{(\Rv/{\rm Mpc}})^3
M_\odot $). Contour levels are indicated by the numbers attached to the curves. }
\label{fig:lensec}
\end{figure*}

The profile (\ref{eq:nfw}) depends on the halo concentration, $c$, and its
virial radius, $\Rv$. 
We determine these two parameters for each halo by matching the combined 
 weak and strong lensing estimates of the projected mass density as given 
 in figure (5) of  Brada{\v c} et el. (2006). 
For convenience, we borrow the results for the projected mass in that figure  and  
show them in figure (\ref{fig:lensem}) here. Points extending to 0.5 and $0.3\rm Mpc$ correspond to the main and sub components,
respectively. Attached to the points are $1\sigma$ error-bars,
taken as $7\%$ and $9\%$ of the mass for the main and the sub-cluster, respectively (Brada{\v c} et el. 2006). 
  We fit NFW profiles to these data by minimizing the quantity
  \begin{equation}
  \tilde \chi=\sum_\alpha {\sigma^{-1}}_\alpha|M^{\rm lens}_\alpha-M^{\rm NFW}_\alpha|
\end{equation} 
with respect to $R_{\rm v}$ and $c$. 
Here the summation is over all data points,  $M^{\rm lens}$ and $M^{\rm NFW}$ are projected masses taken from 
the lensing  measurements and the NFW profile, and $\sigma_\alpha$
are $1\sigma$ error-bars at each data point.   
  This procedure yields  multiple solutions for $\Rv$ and $c$ for 
  both  dark  components.  
 The results are represented in figure (\ref{fig:lensec})
 as contour plot of 
the quantity $\tilde \chi$ as a function of $\Rv$ and $c$
for the main (top panel) and sub (bottom) components. 
There is a clear degeneracy between $\Rv$ and $c$.
The difference between two solutions with similar $\tilde \chi$ is 
illustrated in figure (\ref{fig:lensem}) where,  in addition to the data, 
we  show the projected mass corresponding  to NFW profiles with
  $\Rv=3.2$ \&  2.6 Mpc (for the main) and 
$\Rv=2.9$ \& 2.3 Mpc (for the sub), as indicated in the figure. 
These values of $\Rv$, respectively for the main and the sub,  represent the lower and upper limits of what we consider 
as acceptable match to the mensing measurements. 
Therefore, the largest acceptable ratio between the virial
radii of the main and sub components is $3.2:2.3=1.4:1$.
The mass of a halo within a sphere of radius 
$\Rv$ is
$M_{\rm v}= 10^{14}\left(\frac{h}{0.7}\right)^2\left(\frac{\Rv}{1{\rm Mpc}}\right)^3
M_\odot $. Therefore, the largest mass ratio allowed by these
considerations is $1.4^3:1\approx 3:1$.
We will see in the next section that this ration is also consistent with cosmological initial conditions if the Bullet system resides in a region 
of a mild over-density of 10 times the background density. 

\section{Main-sub dynamics}
\label{sec:eom}

We give here the equations governing the evolution of the 
two dark components of the system IE 0657-56. We neglect the gravity of
gas in the system.
We consider NFW profiles truncated at the virial radii for both components. The system is assumed to be embedded in 
region of density $(1+\delta(t))\bar \rho(t)$ where 
$\bar \rho$ is the mean cosmic matter density and $\delta $ is the density contrast with a time dependence derived from the spherical 
top-hat model. 
We assume a   flat universe with a cosmological constant
$\rhov(t)=const$ to the density from the dark energy. 
 The values of the density parameters are $\rhov/\rho_{\rm crit}=0.7$ 
 and $\bar \rho/\rho_{\rm crit}=0.3$. 
We work in the system of the center of mass of the two clusters. 
The motion of the center of mass, $\vr_{\rm i}$, of the  component $i$ is determined by 
\begin{equation}
\label{fig:eom}
\ddot { \vr}_{\rm i}=\frac{\vF_{\rm i,j}}{M_{\rm i}}-\frac{4\pi}{3}G\left[\bar \rho (1+\delta)-2\rhov\right]{\bf R}_{\rm i}
\end{equation}
where 
 $M_i$ is the mass of each component.   
For a given value of $\delta$ today, the function $\delta(t)$ is obtained from the spherical collapse model.
The force, $\vF_{\rm i,j}$, the halo $j$  applies to 
$i$, is computed for  NFW profiles for both components.
 
For the mass, $\mm$,  of the main component we consider the 
range
 $(1.5-3.4)\times 10^{15} M_\odot  10^{15}$.
This corresponds to 
$\Rv=(2.46- 3.23)\rm Mpc$. According to figure (\ref{fig:lensec}),
$\Rv=2.46 \rm Mpc$ is  ruled out by lensing for all values of the 
concentration parameter, but we explore it anyways and show that such masses are not favored by the timing argument as well. 
For the mass, $\ms$ of the sub, we consider the range 
$ \ms=(0.34-2.8)\times 10^{15} M_\odot$, corresponding to 
$\Rv=(1.5-3)\rm Mpc$. The lower end of this range is inconsistent 
with lensing. 
The concentration parameters, $c$, 
 are chosen to 
be those giving the  best match to the lensing data  (see fig.~\ref{fig:lensec}).

Given a choice of $\mm$ and $\ms$ we numerically solve the 
equations (\ref{fig:eom}) back-in-time for three
values of the current relative velocity, $V$: $4500$, $4000$ and $3700\rm km/s$. 
We assume a distance of $0.7\Mpc$ between the mass centers 
of the two components. 
For simplicity we assume that the relative speed lies along the 
separation between the centers of  mass of the cluster components.

\begin{figure*}
\centering
\begin{sideways}
\mbox{\psfig{figure=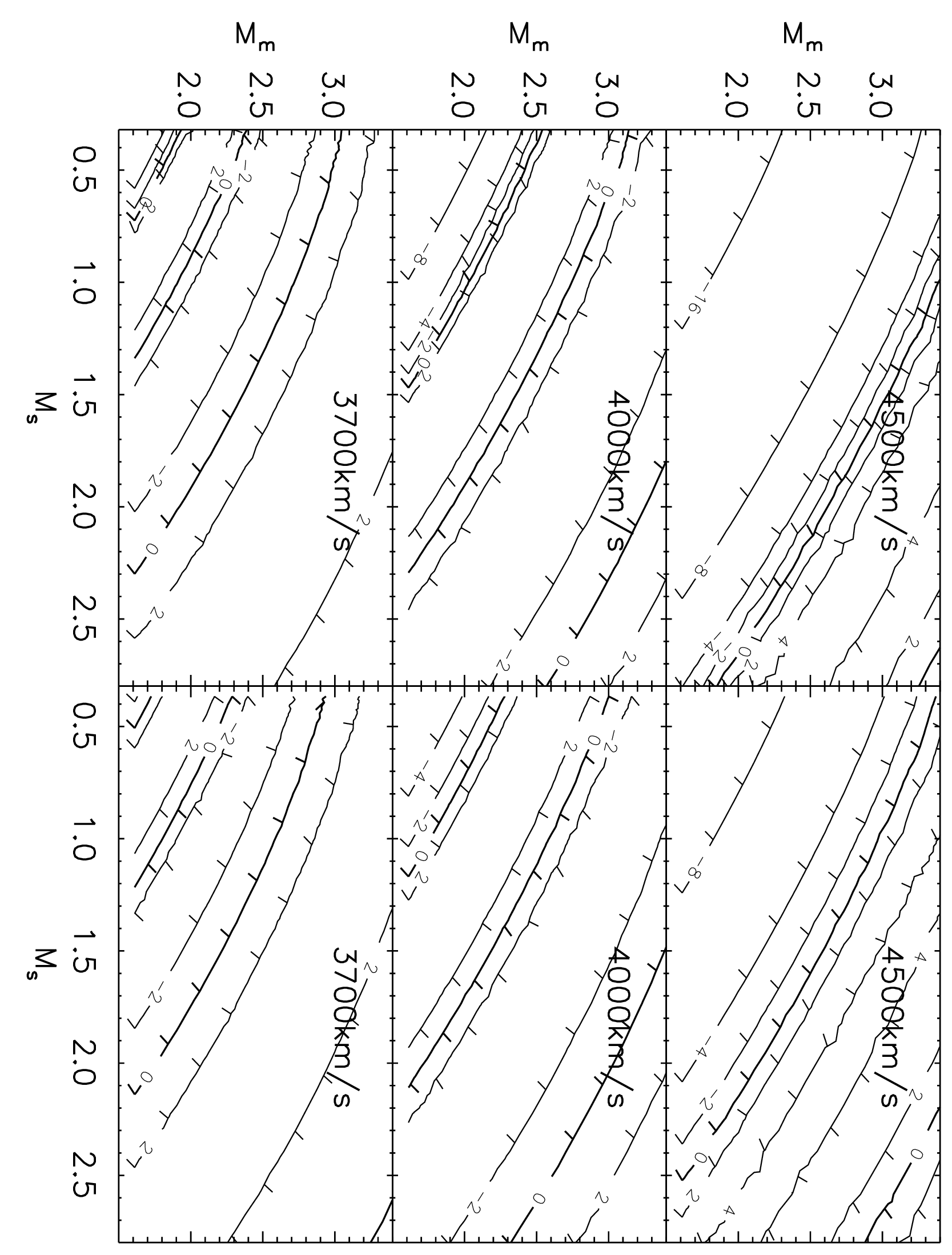,height=6.5in}}
\end{sideways}
\vspace{0.2cm}
\caption{Contours of the separation, $R_{\rm s}-R_{\rm m}$, between the centers of 
mass of the main and sub components (in units of $10^{15} M_\odot$) at $t\rightarrow 0$, as obtained
from  back-in-time solutions of the equations of motion.
 The contours are plotted in the plane of the  assumed masses of the two components.  The contour levels correspond to (physical) separations of -16,-8,-4,-2,0,2,4 and 6 Mpc, as indicated on some of the contours. Thick lines indicate zero separations and ticks indicate downhill directions.
 In each panel in the middle and bottom rows, the relevant set of contours are those lying nearer to 
 the lower masses. For this set, the centers of mass of the two components originate from the same point and reach a maximum separation only once within a Hubble time. 
}
\label{fig:delr}
\end{figure*}

Figure (\ref{fig:delr}) shows the separation, $R_{\rm s}-R_{\rm m}$, 
between the two components at $t\rightarrow 0$, as contour plots in the plane of  $M_{\rm m}$ and $M_{\rm s}$. 
The top, middle and bottom rows, respectively,   $V=4500$, $4000$ and $3700 \rm km/s$. 
The column to the left shows solutions obtained for 
a cluster embedded in mean cosmological density, while the one to the right, to a nonlinear  density contrast of $10$ at the measured redshift of the system. 
Each panel in the middle and bottom rows contains more than one set of contours. 
The set corresponding to  the smallest masses represent the relevant  solutions
in which the separation reaches a maximum value only once within a Hubble time. The separations at $t=0$ are sensitive to the assumed relative 
speed, $V$.
For example, according to the left panels, for   $M_{\rm m}=2\times 10^{15}M_\odot$ 
and $M_{\rm s}=10^{15}M_\odot$, the 
 separation is zero 
for $V=4000 \rm km/s$, while it is $\sim 16 \rm Mpc$ for 
$V=4500\rm km/s$. 
For $V=4500 \rm km/s$

We have seen in the previous section that the lensing analysis 
gives 3:1 as the largest ratio between the mass of the main and the sub-components. 
Inspection of figure  (\ref{fig:delr}) reveals that this is also consistent 
with the requirement of cosmological initial conditions if the 
system is embedded in a region which 10 times denser than the background (top-right panel). The inferred total mass from this figure is $4\times 10^{15}M_\odot=
2.8 {h}^{-1} 10^{15} M_\odot$ which is close to the mass inferred from 
lensing for a 3:1 mass ratio.

The effect of having an over-dense region engulfing the system is most pronounced 
for the largest relative speed (cf the three rows to the right) and is almost 
unnoticeable for the smallest speed. The reason is as follows.
 The existence of this region has little 
effect on the dynamics when the cluster components  are still close to each other.
As we go sufficiently  backward in time, their separation increases 
and so  their mutual gravity decreases.
As this happens,  the density of the region decreases, in accordance with the top-hat model.
For large relative  velocities, large separations are reached while 
the surrounding region is still at a significant over-density and thus can affect 
the dynamics.

\section{Discussion}           
\label{sec:disc}

The relative velocity between the two cluster components is large compared to 
the usual virial speeds ($\sim 10^3 \rm km/s$). 
This triggered discussions on whether or not the system if consistent 
with the currently viable models for structure formation based on the Cold Dark Matter Scenario (CDM) with a cosmological constant, $\Lambda$, i.e.
the $\Lambda\rm CDM$ model. 
Modified Newtonian Dynamics (MOND) has been invoked (Angus, Famaey, \& 
Zhao 2006, Angus \& McGaugh 2007) to explain the estimated relative speed. 
Other explanations relied on long range scalar interactions in the dark sector 
to aid  gravity in boosting the relative speed 
(Farrar \& Rosen 2007). 
Nonetheless, like in the Newtonian case  both of these suggestions need to be
assessed in a cosmological context. Scalar interactions 
modify  large scale  structure formation 
in a desirable way (Nusser, Gubser \& Peebles 2005; Farrar \& Peebles 2004).
Still, proper cosmological simulations need to be performed to assess whether or 
not they could easily reproduce a ``Bullet system".
As for  MOND-like modifications  (Bekenstein 2004), it is still  unclear how 
they can be implemented in a cosmological context. 
The simplest cosmological adaptation had been 
incorporated in a cosmological simulation and lead to large scale structure formation similar 
to $\Lambda \rm CDM$, but with  MOND's acceleration parameter which 
is smaller by an order of magnitude that the standard value (Nusser 2000).

We argue here that even with a large value for the relative speed,
gravitational force alone is consistent with cosmological initial conditions, 
i.e. the timing argument. The total mass we derive
is $2.8 {\rm h}^{-1} 10^{15}M_\odot$, if the system resides in a region 
of a density contrast of about 10.
Hayashi \& White (2006) performed a likelihood analysis 
on the Millennium simulation to conclude that 
the Bullet system can be accommodated in the 
LCDM cosmogony. They adopt masses of $2.16\times 10^{15} h^{-1} M_\odot$ and $5.3 \times 10^{13}h^{-1}M_\odot$ for the main 
and sub components, respectively. 
Our mass estimate is slightly larger than theirs 
but we still could be consistent with the $\Lambda \rm CDM$ model 
normalized to $\sigma_8=0.9$. 
However, the mass ratio we advocate here, 3:1, is quite different from 
the value adopted by Hayashi \& White (2006). 
Zhao (2007) also used the timing argument to constrain the 
mass of the Bullet system without taking into account dynamical effects of an over-dense surrounding environment.
The mutual force in that study is also computed differently from
this work. Zhao's estimate of the total mass is as twice as 
the value derived here.

Springel \& Farrar (2007) used non-cosmological simulations of two colliding 
clusters of a mass ratio of 10:1
 and found that 
the relative speed between the two dark components could be as small as 
2700 km/s, while the shock speed is 4500 km/s. 
As mentioned before, a 10:1 mass ratio is inconsistent with lensing. 
Of course, the physical effects leading to velocity lag of dark 
mass relative to the shock could still be important   and  they need to be quantified with
simulations having a mass ration of 3:1.
Nonetheless,  for low relative speeds the timing argument yields  masses  
  (see bottom-left panel in fig.~\ref{fig:delr}) which
  are too small to be consistent with the lensing measurements.

We have made some attempts, the details of which are not presented here, at including dynamical friction and tidal stripping according to the recipes in Nusser \& Sheth (1999). 
These effects reduce the relative speed between the two
components, therefore, 
for a given final relative speed they increase corresponding masses required to match the timing constraint.

The analysis presented here  neglects
important processes which might affect the dynamics of the system. 
Each of the two components have formed by merging of smaller halos. 
Therefore, sometime in the past, the matter making each component
was distributed in more than one clump. 
At that time, the mutual force determining the relative motion of the centers of mass of the two components should depend on the spatial distribution of their progenitors.
If, however, major merging and accretion activities in both 
components ceased when the separation between them was large enough. Then
the monopole term alone, which we use here for the mutual force, should be a reasonable description to 
the dynamics of the system.  

We have tuned the masses so that the solution back-in-time
gives vanishing separation near $t=0$. This has been done by solving 
the initial value problem of given 
current separation and relative speed as input to the numerical solution.
Alternatively, we could have solved a boundary value problem where 
the first boundary condition is   
current separation and the second is a constraint which guarantees 
the cosmological constraint near $t=0$ (e.g. Peebles 1993, Nusser \& Branchini).
This could be done using the least action principle, as proposed
by Peebles (1989). This approach yields a prediction for the 
current relative velocity for a given mass choice. This predicted velocity 
could then be compared with the observed speed.
This approach will be employed in future work.

\section*{acknowledgment}
This research is supported by the German-Israeli Foundation for 
Development and Research and the Asher Space Research Fund. 
 
\protect



\end{document}